\documentclass[prd,amsmath,amssymb,preprint]{revtex4}
\usepackage{graphicx}
\usepackage[english]{babel}
\usepackage{dcolumn}
\usepackage{bm}
\usepackage{longtable}
\usepackage{pdflscape}
\usepackage{hyperref}
\usepackage{amsmath}
\renewcommand{\vec}[1]{{\mbox{\boldmath$#1$}}}

\begin{document}
\title{Isotope shifts of the $1s^22s2p(J)$ -$1s^22s^2$ transition energies in Be-like thorium and uranium}
\author{N.~A.~Zubova$^{1,2}$, I.~S.~Anisimova$^1$, M.~Yu.~Kaygorodov$^1$, Yu.~S.~Kozhedub$^{1,2}$, A.~V.~Malyshev$^{1}$, V.~M.~Shabaev$^{1}$, I.~ I.~Tupitsyn$^{1,3}$, G.~Plunien$^{4}$, C.~Brandau$^{5,6,7}$, and Th.~St\"ohlker$^{5,8,9}$}
\address{$^1$Department of Physics, St. Petersburg State University,
7/9 Universitetskaya nab., St.~Petersburg 199034, Russia \\
$^2$NRC “Kurchatov Institute”, Academician Kurchatov 1, Moscow 123182, Russia
\\
$^3$Center for Advanced Studies, Peter the Great St. Petersburg Polytechnic
University, Polytekhnicheskaja 29, St. Petersburg  195251, Russia\\
$^4$Institut f\"ur Theoretische Physik, TU Dresden, Mommsenstrasse 13, Dresden,
D-01062, Germany\\
$^5$GSI Helmholtzzentrum f\"ur Schwerionenforschung GmbH, D-64291 
Darmstadt, Germany\\
$^6$ExtreMe Matter Institute EMMI and Research Division, GSI 
Helmholtzzentrum f\"ur Schwerionenforschung, D-64291 Darmstadt, Germany\\
$^7$Institut f\"ur Atom- und Molek\"ulphysik, Justus-Liebig-University Giessen, Leihgesterner Weg 217, D-35392 Giessen, Germany\\
$^8$Helmholtz-Institut Jena, D-07743 Jena, Germany\\
$^9$Institut f\"ur Optik und Quantenelektronik, 
Friedrich-Schiller-Universit\"at Jena, D-07743 Jena, Germany}
\begin{abstract}
  Precise calculations of the isotope shifts in berylliumlike
  thorium and uranium ions are presented. The main contributions to the
  field and mass shifts are
  calculated within the framework of the Dirac-Coulomb-Breit Hamiltonian
  employing the configuration-interaction Dirac-Fock-Sturm method. These
  calculations include the relativistic, electron-electron correlation,
  and Breit-interaction effects.
  The QED, nuclear deformation, and nuclear polarization corrections
  are also evaluated. 
\end{abstract}
\maketitle
\section{Introduction}
Theoretical and experimental studies of the isotope shifts in highly charged ions
can provide tests of the relativistic and QED theory of the nuclear recoil effect
in nonperturbative regime and serve as a good tool for precise determination of the
nuclear charge radius differences.
First isotope shift measurements in highly charged ions were performed in Refs. \cite{ell96,ell98,sch05}. 
The measurements of the isotope shifts of the binding energies in B-like argon \cite{cre06}
and in Li-like neodymium \cite{Brandau_2008} have provided the first tests of the relativistic theory
of the nuclear recoil effect with highly charged ions. The latter experiment led also to determination
of the nuclear charge radius difference for the $^{142,150}$Nd  isotopes. The use of the dielectronic recombination technique \cite{Brandau_2008,bra10,bra13} at the GSI/FAIR facilities allows also the related experiments for heavy Be-like ions.

The main goal of the present work is to extend our calculations of the isotope shifts
in Li- and B-like ions \cite{Zubova_2014,Zubova_2016} to Be-like thorium and uranium ions,
which are of most interest for the experimental study. Previously, the calculations of the
isotope shifts in Be-like ions have been performed in the range $Z=5-74$ with the use of the multiconfiguration  Dirac-Fock method \cite{Naze_2014}. The calculations of the transition energies in these ions have been performed in Refs. \cite{Cheng_2018,Kaigorodov_2019} (see also Ref. \cite{Yerokhin_2015,Yerokhin_2014} and references therein).  

The precision of the isotope shift measurements in heavy ions is approaching  the level of the QED effects.
Moreover, it is expected that at the FAIR facilities this precision will be improved by an order of magnitude.
It means that the relevant theoretical calculations must be performed to the utmost accuracy.
In the present paper, the dominant contributions to the isotope shifts are calculated employing the Dirac-Coulomb-Breit Hamiltonian. These calculations, which are based on the
Dirac-Fock-Sturm method \cite{tup03,Tupitsyn_2018},
include the relativistic, electron-electron correlation, and Breit-interaction
effects. Additionaly, we evaluate the QED, nuclear deformation, and nuclear polarization corrections
which become rather large for heavy ions. 

The relativistic units ($\hbar=c=1$) are used throughout the paper.
\section{Theory}
\subsection{Nuclear size effect}
The finite nuclear size effect (the so-called field shift) is caused by the difference in the nuclear charge distribution
of the isotopes. The main contribution to the field shift can be calculated in the framework of the
Dirac-Coulomb-Breit Hamiltonian. The nuclear charge distribution is usually approximated by the
spherically-symmetric Fermi model
\begin{equation}
\label{rho}
\rho(r,R)=\frac{N}{1+{\rm{exp}}[(r-c)/a]},
\end{equation}
where the parameter $a$ is generally fixed to be $a=2.3/(4{\rm \ln}3)$ fm and the parameters $N$ and $c$ are determined using the given value of the root-mean-square ($\rm{rms}$) nuclear charge radius $R=\langle r^2 \rangle^{1/2}$ and the normalization condition $\int{d\vec{r} \rho({r},R)}=1$. 
The potential induced by the nuclear charge distribution $\rho(r,R)$ is defined as 
\begin{equation}
\label{Vn}
V_{N}(r,R)= -4\pi \alpha Z \int\limits_{0}^{\infty} {dr' r'^2 \rho (r',R) \frac{1}{r_{>}}},
\end{equation}
where $r_{>}={\rm{{max}}}(r,r')$. 
Since the finite nuclear size effect is mainly determined by the $\rm{rms}$ nuclear charge radius,
the energy difference between two isotopes can be approximated as
\begin{equation}
\label{FS_1}
\delta E_{FS} = {F}\delta \langle r^2 \rangle,
\end{equation}
where $F$ is the field shift factor and $\delta \langle r^2 \rangle$ is the 
mean-square charge radius difference. 
In accordance with this definition and the virial theorem, the $F$ factor can be also evaluated by
\begin{equation}
\label{FS_2}
F= \langle \psi \mid \sum_{i} \frac{dV_{N}(r_{i},R)}{d\langle r^2 \rangle} \mid  \psi \rangle,
\end{equation}
where $\psi$ is the wave function of the state under consideration and the index $i$ runs over all atomic electrons.  

\subsection{Relativistic nuclear recoil effect}
The fully relativistic theory of the nuclear recoil effect can be formulated only in the framework of quantum
electrodynamics \cite{Shabaev_1985,Shabaev_1988,pac95,sha98,adk07}.
However, to the lowest relativistic order (within the Breit approximation),
the nuclear recoil effect can be taken into account
using the effective recoil operator \cite{Shabaev_1985,Shabaev_1988,Palmer_1987}:
\begin{eqnarray}
\label{HM}
H_M &=& \frac{1}{2M}\sum_{i,k}\Bigl[
{\vec{p_i}}\cdot {\vec{p_k}} -\frac{\alpha Z}{r_i}\Bigl[\vec{\alpha_i}+
\frac{(\vec{\alpha_i}\cdot\vec{r_i})\vec{r_i}}{r_i^2}\Bigr]\cdot\vec{p_k}
\Bigr] \,.
\end{eqnarray}
This operator can be used for relativistic calculations of the nuclear recoil effect
in ions and atoms (see, e.g., Refs. \cite{Zubova_2014,Zubova_2016,Naze_2014,tup03,Tupitsyn_2018,fis16,fil17} and references therein).
The calculation is carried out by averaging the operator (\ref{HM})
with the eigenvectors of the Dirac-Coulomb-Breit Hamiltonian.

\subsection{QED, nuclear deformation and nuclear polarization corrections}
Since our consideration is restricted to very heavy ions, the independent-electron approximation
can be used to evaluate the QED, nuclear deformation, and nuclear polarization corrections.

To calculate the self-energy and vacuum-polarization corrections to the field shift,
one can use analytical formulas for these corrections derived for H-like ions in Ref.
\cite{Milstein_2004}. In case of uranium, an approximate formula obtained
by fitting the direct numerical calculations \cite{Yerokhin_2011} can be also employed.

The QED calculation of the one-electron recoil effect for $n=1,2$ states was performed in Refs.
\cite{art95a,sha98b,adk07,mal18}. In addition, two-electron recoil contributions of zeroth order in $1/Z$ should be taken into account for the $1s^22s2p(J)$ states. 
A detailed analysis of the relevant contributions for He-like ions was presented in Ref. \cite{mal18}.

To evaluate the nuclear deformation effect, one has to replace the standard spherically symmetric
Fermi model for the nuclear charge distrubution by \cite{koz08} 
\begin{eqnarray} \label{rho_def}
\rho(r) = \frac{1}{4\pi}\int{d\vec{n} \rho(\vec{r})}\,,
\end{eqnarray}
where { $\rho(\vec{r})$}  is the axially-symmetric Fermi distribution,
\begin{equation}
\label{rho1}
\rho(\vec{r})=\frac{N}{1+{\rm{exp}}[(r-r_0(1+\beta_{20}Y_{20}(\theta)+ \beta_{40}Y_{40}(\theta))/a]}\,,
\end{equation}
consistent  with the normalization condition
{$\int{d\vec{r} \rho(\vec{r})}=1$}.
Here $Y_{20}(\theta)$ and $Y_{40}(\theta)$ are  the spherical functions,
$\beta_{20}$ and $\beta_{40}$ are the quadrupole and hexadecapole deformation parameters
\cite{koz08,bem73,zum84,moe95}.
The difference between the nuclear size effect obtained with the deformed model (\ref{rho_def})
and the standard spherically-symmetric
Fermi model (\ref{rho})
at the same rms radius is ascribed to the nuclear deformation effect.

Finally, the interaction between the electrons and the nucleons causes the
nucleus to make virtual transitions to excited states.
This results in the increase of the binding energy of the electrons. 
To evaluate this effect, which is known as the nuclear polarization effect,
one should consider the two-photon electron-nucleus interaction diagrams in which the
intermediate nuclear states are excited. The calculations of this effect
were performed in Refs. \cite{plu95,plu96,nef96,vol14}.

\section{Results and discussion}
In Tables \ref{tabIS1}, \ref{tabIS2} the individual contributions to the isotope shifts of
the $1s^22s2p(J) - 1s^22s^2$ transition energies in $^{232,230}$Th$^{86+}$, $^{238,236}$U$^{88+}$,
and $^{238,234}$U$^{88+}$ are presented. The nuclear charge radii and the $\delta \langle r^2 \rangle $
differences  have been taken from Ref. \cite{ang13}.
The field shifts are evaluated within the framework of the Dirac-Coulomb-Breit Hamiltonian using the formulas
(\ref{FS_1})-(\ref{FS_2}).
The calculations are performed using the conguration-interaction Dirac-Fock-Sturm method for an extended nucleus \cite{tup03}.
The excited configurations are obtained from the basic configuration via single, double, and triple excitations
of electrons.  
The accuracy of the
calculations is defined by the stability of the results with respect to a variation of the basis size.
The same method has been used to calculate the mass shifts within the approximation defined
by the effective nuclear recoil operator (\ref {HM}).

The QED corrections to the field shifts have been evaluted in the one-electron approximation
using the related formulas from Ref. \cite{Milstein_2004,Yerokhin_2011}. The QED effect on the mass shift
was obtained as a sum of one- and two-electron contributions evaluated to zeroth order in $1/Z$.
The one-electron terms have been taken from Ref. \cite{mal18} while the two-electron
corrections have been calculated in this work.
As one can see from the tables,  the QED recoil contribution  is even larger than the mass
shift obtained within the framework of the Breit approximation. 
The nuclear deformation and polarization effects have been taken from the related calculations
for Li-like thorium and uranium \cite{Zubova_2014}. The uncertainties of these effects
determine the total theoretical uncertainties.
\begin{table}
\small
\caption{\label{tabIS1}
Individual contributions to the isotope shifts of the
$1s^22s2p_{1/2}(J) - 1s^22s^2$ transition energies in $^{232,230}$Th$^{86+}$,
$^{238,236}$U$^{88+}$, and $^{238,234}$U$^{88+}$ (in meV) with given values of $\delta \langle r^2 \rangle $.} 
\begin{tabular}{c|c|c|c}
\hline
& $^{232,230}$Th$^{86+}$ & $^{238,236}$U$^{88+}$ & $^{238,234}$U$^{88+}$\\
& $^{232,230}\delta \langle r^2 \rangle$=0.2050 $\rm{fm^2}$ &
$^{238,236}\delta \langle r^2 \rangle$=0.1676 $\rm{fm^2}$ & 
$^{238,234}\delta \langle r^2 \rangle$=0.334 $\rm{fm^2}$\\
\hline
&\multicolumn{3}{c}{$1s^22s2p_{1/2}(J=0) - 1s^22s^2$}\\
\hline
Main contributions & &  &\\
Field shift &$-$112.4 & $-$110.8 &$-$220.7 \\
Mass shift & 0.1&0.1 & 0.2\\
QED & & &\\
Field shift & 0.6&0.6&1.2\\
Mass shift & 0.4&0.4 & 0.9\\
Others&& &\\
Nuclear polarization &1.6 &1.1 & 2.3 \\
Nuclear deformation & 1.5&$-$2.2 &$-$2.4 \\
Total IS theory$^{a}$ & $-$108.2(22) & $-$110.8(31) &$-$218.5(32)\\
\hline
&\multicolumn{3}{c}{$1s^22s2p_{1/2}(J=1) - 1s^22s^2$}\\
\hline
Main contributions & & &\\
Field shift &$-$112.8& $-$111.0 & $-$221.3\\
Mass shift & 0.1 & 0.1 & 0.2 \\
QED &&&\\
Field shift &0.6&0.6&1.2\\
Mass shift &0.4&0.4 &0.9\\
Others&& &\\
Nuclear polarization &1.6 &1.1& 2.3 \\
Nuclear deformation & 1.5&$-$2.2  & $-$2.4\\
Total IS theory$^{a}$  & $-$108.5(22) &$-$111.1(31) & $-$219.1(32)\\
\hline
\hline
\end{tabular}
$^{a}${The uncertainty of  $\delta{\langle r^2 \rangle}$ is not included.}
\end{table}
\begin{table}
\small
\caption{\label{tabIS2}
Individual contributions to the isotope shifts of the
$1s^22s2p_{3/2}(J) - 1s^22s^2$ transition energies in $^{232,230}$Th$^{86+}$,
$^{238,236}$U$^{88+}$ and $^{238,234}$U$^{88+}$ (in meV)
with given values of $\delta \langle r^2 \rangle$.} 
\begin{tabular}{c|c|c|c}
\hline
& $^{232,230}$Th$^{86+}$ & $^{238,236}$U$^{88+}$ & $^{238,234}$U$^{88+}$\\
& $^{232,230}\delta \langle r^2 \rangle$=0.2050 $\rm{fm^2}$ &
$^{238,236}\delta \langle r^2 \rangle$=0.1676 $\rm{fm^2}$ & 
$^{238,234}\delta \langle r^2 \rangle$=0.334 $\rm{fm^2}$\\
\hline
&\multicolumn{3}{c}{$1s^22s2p_{3/2}(J=2) - 1s^22s^2$}\\
\hline
Main contributions & & &\\
Field shift & $-$124.6 &$-$123.7 &$-$246.4\\
Mass shift & 0.3 & 0.3 &0.6\\
QED & &\\
Field shift &0.9 & 0.9 & 1.8 \\
Mass shift & 0.4 & 0.4 & 0.8\\
Others&&\\
Nuclear polarization &1.7 &1.2 &2.6\\
Nuclear deformation & 1.5 &$-$2.4 & $-$2.7\\
Total IS theory$^{a}$  & $-$119.8(22) & $-$123.2(32) &$-$243.2(33)\\
\hline
&\multicolumn{3}{c}{$1s^22s2p_{3/2}(J=1) - 1s^22s^2$}\\
\hline
Main contributions & & &\\
Field shift &$-$125.2 &$-$124.3 &$-$247.7\\
Mass shift & 0.3 & 0.3 & 0.6\\
QED & & &\\
Field shift &0.9 & 0.9&1.8 \\
Mass shift & 0.4 & 0.4 & 0.8\\
Others&& &\\
Nuclear polarization &1.7 &1.2 & 2.6\\
Nuclear deformation &1.5 &$-$2.4 & $-$2.7\\
Total IS theory$^{a}$& $-$120.4(22) &$-$123.9(32) &$-$244.5(33)\\
\hline
\hline
\end{tabular}
$^{a}${The uncertainty of  $\delta{\langle r^2 \rangle}$ is not included.}
\end{table}
\section{Conclusion}
The isotope shifts of the $1s^2 2s2p(J) - 1s^2 2s^2$ transition energies
 in Be-like  thorium and uranium ions are calculated 
including the relativistic, electron-electron correlation, Breit, and QED
contributions.  The nuclear polarization and nuclear deformation  
corrections are taken into account within the framework of the independent-electron approximation.
The QED effects contribute on the level of the total uncertainty which is  mainly defined by the
nuclear polarization and deformation effects.

\begin{acknowledgments}
  This work was supported by RFBR (Grant No.~18-32-00275),
  SPSU-DFG (Grants No. 11.65.41.2017 and No. STO 346/5-1), Ministry of Education and Science of the Russian Federation
Grant No. 3.1463.2017/4.6, and DAAD Programm Ostpartnerschaften, TU Dresden.
  M.Y.K., A.V.M., and V.M.S.  acknowledge also support from
  the Foundation for the advancement of theoretical physics and
mathematics ``BASIS''. 
\end{acknowledgments}
\clearpage

\end{document}